\documentclass[prl, aps, twocolumn, superscriptaddress]{revtex4-1}
\usepackage{graphicx}
\usepackage[english]{babel}
\usepackage{amsmath}
\usepackage{amssymb}
\usepackage{tensor}
\usepackage{xcolor}
\usepackage[percent]{overpic}       

\newcommand{\ba}{\begin{eqnarray}}
    \newcommand{\ea}{\end{eqnarray}}

\usepackage[colorlinks,citecolor=blue,linkcolor=blue,breaklinks=true]{hyperref}
\begin{document}

    \title{Dissipative Spectroscopy}
    
    \author{Xudong He}
    \affiliation{Graduate School of China Academy of Engineering Physics, Beijing 100193, China}
    
    \author{Yu Chen}
    \email{ychen@gscaep.ac.cn}
    \affiliation{Graduate School of China Academy of Engineering Physics, Beijing 100193, China}
    \date{\today}
    
    \begin{abstract}
        We introduce dissipative spectroscopy as a framework for extracting spectral information from quantum systems via controlled dissipation. By establishing a general dissipative response theory applicable to both Markovian and non‑Markovian environments, we develop a protocol to access the dissipative spectrum (DS) through driven oscillation‑dissipation resonance. We show that the DS can identify two‑particle soft modes near quantum critical points and, on the normal‑phase side, predict the emergence of macroscopic order exhibiting power‑law growth following a dissipation quench. These distinctive signatures appear in quasiparticle‑dominant regimes, previously considered trivial. Furthermore, we introduce extended dissipative susceptibilities that capture leading memory effects and demonstrate their utility in a dissipative fermionic model. Our results indicate that the DS is readily accessible and offers a versatile tool for probing equilibrium properties as well as predicting nonequilibrium dissipative dynamics.
    \end{abstract}

    \maketitle
    
   Spectroscopy stands as a cornerstone of modern physics, where linear response theory has long served as the principal framework for decoding quantum correlations via external perturbations \cite{kubo1957,kubo1966,mahan2000}. Established techniques—such as angle-resolved photoemission spectroscopy \cite{shen2021}, neutron scattering \cite{lovesey1984,dai2015}, and optical conductivity measurements \cite{Haule11,dressel2002}—are rooted in this Hermitian paradigm, linking equilibrium correlation functions to dynamical susceptibilities and revealing phenomena from quasiparticle dispersions to collective excitations. When quantum detections are nowadays focusing on weak measurements aiming at coherent extraction of information \cite{WM}, the influence of external force and noise fluctuations become comparable. Equipped by recent developments in dissipation engineering \cite{Cirac09,Zoller08,Zoller12,Mueller22,Diehl16,Daley14,DissIon11,Cooper20,Cooper21,Cooper24,Pekker19a,Pekker19b,Kollath23,Dieter16,Jiang16, Kamenev23, Diehl25, Weimer21}, we are ready to distinguish the contributions from these different sources. A quest is then raised: can one formulate a spectroscopy that exploits noise-driven dissipative processes rather than conventional external forces, and what distinct physical insights might such an approach yield?

Recent progresses in monitoring open quantum dynamics suggests the possibility of probing nonequilibrium behavior directly through dissipation \cite{CH20,Chen21,ENS20,ENS23,Wenlan23}. By treating environmental coupling as a controlled perturbation, a non-Hermitian linear response theory (NHLRT) has been developed, providing a generalized susceptibility that encodes spectral information via early-time dissipative transients. This formalism has shed light on entropy dynamics \cite{Chen21}, yielded generalized \(f\)-sum rules in dissipative settings \cite{Zhang24}, and revealed connections between dissipative responses and anomalous dimensions in strongly correlated quantum gases \cite{Wenlan23}. Despite these advances, the spectroscopic implications of dissipative response remain largely unexplored, with most studies focused on real-time evolution rather than spectral reconstruction.

In this Letter, we introduce a general framework for dissipative spectroscopy applicable to both Markovian and non-Markovian environments. We design a dynamical protocol that extracts spectral information through a controlled dissipative probe and establish that the resulting dissipative spectrum (DS) uncovers soft modes near quantum critical points. Strikingly, even when the DS exhibits Lorentzian peaks—indicating well-defined quasiparticle excitations—their scaling behavior signals the emergence of macroscopic order on the disordered side of the transition, a phenomenon absent in unitary quenches confined to the normal phase. Thus, DS exposes a distinct form of dissipation-induced quantum criticality. We further show how memory effects can be resolved via generalized dissipative susceptibilities and clarify the conditions under which DS remains valid. Our work find an alternative pathway for spectral diagnostics in open quantum settings, extending beyond the reach of conventional Hermitian spectroscopy.
    
    \emph{\color{blue}General Dissipative Response Theory and Dissipative Spectroscopy\color{black}}. --- Here we are going to present a general dissipative response theory (DRT) where series of generalized susceptibilities and spectroscopies can be defined. We study an open quantum system described by the Hamiltonian $\hat{H} = \hat{H}_{\rm S} + \hat{H}_{\rm E} + \hat{V}_{\rm SE}$, where the system–environment (SE) coupling takes the form $\hat{V}_{\rm SE} = \eta \sum_j (\hat{O}_j \hat{\xi}_j^\dagger + \hat{O}_j^\dagger \hat{\xi}_j)$. $\hat{H}_{\rm S}$ ($\hat{H}_{\rm E}$) denotes the system (environment) Hamiltonian, $\hat{O}_j$ ($\hat{\xi}_j$) is a system (environment) operator, $j$ is the index of the operator, and $\eta$ the coupling strength. The environment is modeled as a Gaussian bath, so that all higher-order correlation functions factorize into products of two-point functions. We assume $\langle \hat{\xi}^{}_j(t)\rangle_{\rm E} = \langle \hat{\xi}^\dagger_j(t)\rangle_{\rm E} = 0$, $\langle \hat{\xi}^{}_i(t)\hat{\xi}_j^\dagger (t')\rangle_{\rm E} = \delta_{ij}g^>_\xi(t-t')$, and $\langle \hat{\xi}_i^\dagger(t)\hat{\xi}^{}_j(t')\rangle_{\rm E} = \delta_{ij}g^<_\xi(t-t')$, with $\langle \cdots \rangle_{\rm E}$ representing ensemble average on the enviroment.
    
    Treating $\hat{V}_{\rm SE}$ as a perturbation in the interaction picture, the deviation of a physical observable $\hat{W}$ from its unperturbed value is
    \ba
    \delta\mathcal{W}(t)= \left\langle \hat{\mathcal{U}}^{I,\dagger}(t) \hat{W}(t) \hat{\mathcal{U}}^I(t) - \hat{W}(t) \right\rangle_{\rm S,E},
    \ea
    where $\hat{\mathcal{U}}^I(t) = \mathcal{T}_t \exp\left(-i\int_0^t \hat{V}_{\rm SE}(t') dt'\right)$ is the evolution operator in the interaction picture, $\hat{W}(t)$ is in the Heisenberg picture with respect to $\hat{H}_{\rm S}$, and $\langle \cdots \rangle_{\rm S,E}$ denotes the ensemble average over the initial product state of system and environment. Expanding to second order in $\hat{V}_{\rm SE}$ yields the DRT:
    \ba
    \delta\mathcal{W}(t)&=& \eta^2\sum_j \!\iint_0^t \!\!\!dt_1 dt_2 \,  \left[ \left\langle \mathcal{L}_{\hat{O}_j^\dagger(t_1),\hat{O}_j(t_2)}\hat{W}(t) \right\rangle_{\rm S} g^>_\xi(t_{12}) \right.\nonumber\\
    && \left. +\left\langle \mathcal{L}_{\hat{O}_j(t_1),\hat{O}_j^\dagger(t_2)}\hat{W}(t) \right\rangle_{\rm S} g^<_\xi(t_{12}) \right] ,\label{GDRT}
    \ea
    with \(\langle\cdots\rangle_{\rm S}\) denotes ensemble average over the initial state of system, and the superoperator $\mathcal{L}$ defined as
    \ba
    \mathcal{L}_{\hat{A}(t_1),\hat{B}(t_2)}\hat{W}(t) \!&\equiv&\! \hat{A}(t_1)\hat{W}(t)\hat{B}(t_2)\!-\! \theta_{12}\hat{W}(t)\hat{A}(t_1\!)\hat{B}(t_2\!)\nonumber\\
    &&- \theta_{21} \hat{A}(t_1)\hat{B}(t_2)\hat{W}(t),
    \ea
    and $\theta_{ij}=\theta(t_{ij})$ the Heaviside step function, $t_{ij}=t_i-t_j$.
    
    In the low-temperature Markovian limit, where $\eta^2 g^>_\xi(t-t') = 2\gamma\delta(t-t')$ and $g^<_\xi(t-t') = 0$ ($\gamma$ being the dissipation strength), the DRT reduces to the NHLRT:
    \ba
    \delta\mathcal{W}(t)=  \gamma\int_0^t dt' \sum_j \chi^{\mathcal{D}}_{\hat{O}_{j},\hat{W}}(t-t'),
    \ea
    where $\chi^{\mathcal{D}}_{\hat{O}_{j},\hat{W}}(t-t') = 2\left\langle \mathcal{L}_{\hat{O}_{j}^\dagger(t'),\hat{O}^{}_{j}(t')} \hat{W}(t) \right\rangle_{\rm S}$ is the dissipative susceptibility. For brevity, $\chi_{\hat{O},\hat{W}}^{\cal D}=\sum_j\chi_{\hat{O}_j,\hat{W}}^{\cal D}$.
    
    Now suppose the strength of $V_{\rm SE}$ is modulated as $\gamma(t) = \gamma + \gamma'\cos(\omega_0 t)$, with $\gamma' \ll \gamma$. The leading dissipative response reads \(\delta\mathcal{W}(t) = \int_0^t dt'  \left( \gamma\! +\! \gamma'\cos(\omega_0 t')\right) \chi^{\mathcal{D}}_{\hat{O},\hat{W}}(t\!-\!t').\) Subtracting the constant-$\gamma$ background leaves the oscillatory part
    \[
    \delta\mathcal{W}_{\rm osc}(t) = \gamma' \int_0^t \chi^\mathcal{D}_{\hat{O},\hat{W}}(t-t') \cos(\omega_0 t') \, dt'.
    \]
    When $\omega_0^{-1} \ll t \ll \gamma^{-1}$, $\delta{\cal W}_{\rm osc}(t)$ is dominated by the resonant component:
    \ba
    \delta\mathcal{W}_{\rm osc}(t) \!\approx \!\gamma' \left| \chi^{\cal D}_{\hat{O},\hat{W}}(\omega_0) \right|\!t \cos\left( \omega_0 t \!+\!\arg \chi^{\cal D}_{\hat{O},\hat{W}}(\omega_0) \right)\!\!,
    \ea
    where $\chi_{\hat{O},\hat{W}}^{\cal D}(\omega)$ is the DS, defined as the Fourier transformation of the dissipative susceptibility. The linear growth in $t$ reflects resonance. Both the amplitude and phase of this signal allow reconstruction of the DS. Although the present scheme requires $\omega_0^{-1}\ll\gamma^{-1}$, protocols can be extended for relative large $\gamma$ \cite{supplementary}.
    
    In practice, the Markovian limit is rarely exact. Accessing the DS therefore requires knowledge of the memory response, which remains within the present framework. Introducing $\bar{t}=\frac{1}{2}(t_1+t_2)$ and $\delta t=t_{12}$, and expanding in $\delta t$ \cite{supplementary}, Eq.~\eqref{GDRT} can be written as
    \ba
    \delta\mathcal{W}(t) &\approx& \eta^2  \sum_{j,\ell} \left[ \chi^{\mathcal{D},[\ell]\Re}_{\hat{O}_j,\hat{W}}\circ\mathcal{G}^{>,[\ell]\Re}_\xi +\chi^{\mathcal{D},[\ell]\Im}_{\hat{O}_j,\hat{W}}\circ\mathcal{G}^{>,[\ell]\Im}_\xi \right.\nonumber\\
    &&\left. \chi^{\mathcal{D},[\ell]\Re}_{\hat{O}^{\dagger}_j,\hat{W}}\circ \mathcal{G}^{<,[\ell]\Re}_\xi+\chi^{\mathcal{D},[\ell]\Im}_{\hat{O}^{\dagger}_j,\hat{W}}\circ \mathcal{G}^{<,[\ell]\Im}_\xi\right](t,0),
    \ea
    where $\circ$ denotes the convolution $(A\circ B)(t,0)\equiv \int_0^t d\bar{t}\,A(t-\bar{t})B(\bar{t}-0)$, $\ell$ is the expansion order in $\delta t$.
    \ba
    \mathcal{G}^{\gtrless,[\ell]\Re/\Im}_\xi(\bar{t}) = \int_0^{t-|t-2\bar{t}|} \!\!\!\!\!d\delta t \ \Re/\Im\!\left(g^{\gtrless}_\xi(\delta t)\right)\frac{\delta t^\ell}{\ell! 2^\ell},
    \label{def_g}
    \ea
    with $\Re/\Im(\#)$ taking the real or imaginary part of $\#$. The dissipative susceptibilities \(\chi^{\mathcal{D},[\ell]\Re/\Im}_{\hat{A},\hat{W}}\) are
    \ba
    \chi^{\mathcal{D},[0]\Re}_{\hat{A},\hat{W}}\! &=&\! 2\left\langle \mathcal{L}_{\hat{A}^\dagger,\hat{A}}\hat{W}\right\rangle_{\rm S}, \ 
    \chi^{\mathcal{D},[0]\Im}_{\hat{A},\hat{W}} \!=\! i \left\langle \left[\hat{A}^{\dagger}\hat{A}, \hat{W} \right] \right\rangle_{\rm S},
    \label{def_chi_1}\nonumber\\
    \chi^{\mathcal{D},[1]\Re}_{\hat{A},\hat{W}}&=& i\left\langle \left[[\hat{H}_{\rm S},\hat{A}^\dagger]\hat{A}, \hat{W}\right] - \left[\hat{A}^\dagger[\hat{H}_{\rm S},\hat{A}], \hat{W}\right] \right\rangle_{\rm S}\!,
    \label{def_chi_2}\\
    \chi^{\mathcal{D},[1]\Im}_{\hat{A},\hat{W}} &=& 2\left[ -\left\langle \mathcal{L}_{[\hat{H}_{\rm S},\hat{A}^\dagger],\hat{A}}\hat{W}+ \mathcal{L}_{\hat{A}^\dagger,[\hat{H}_{\rm S},\hat{A}]}\hat{W} \right\rangle_{\rm S}\right],
    \label{def_chi_3}\nonumber
    \ea
    where $\hat{A}$ denotes a generic operator. Time is omitted above and can be restored by replacing $\hat{W}$ with $\hat{W}(t)$.
    
    We note that ${\cal G}_\xi^{\gtrless,[\ell]} \propto \tau_0^\ell$, where $\tau_0$ is the correlation time of $g_\xi(t)$, indicting memory timescale. Similarly, $\chi^{{\cal D},[\ell]} \propto \Lambda^\ell$, with $\Lambda$ a characteristic energy scale of the system. The expansion is therefore controlled by $\Lambda \tau_0 $. We thus find a truncation of $\ell\leq 1$ being enough for describing the main memory effects for small $\Lambda\tau_0$.
    
    Below we present three illustrative applications: (1) dynamical measurement of the DS in a free-fermion chain; (2) critical dissipative dynamics near critical point; and (3) a direct demonstration of memory effects captured by dissipation within the $\ell\leq 1$ approximation.
    
    \emph{\color{blue}Example for measuring DS: dissipative free fermions\color{black}}. --- 
    We demonstrate a dynamical measurement scheme using a dissipative tight-binding chain of spinless fermions:
    \ba
    \hat{H}_{\rm FF} = -h_0 \sum_j (\hat{c}_j^\dagger \hat{c}_{j+1} + \text{H.c.}) - \mu \sum_j \hat{c}_j^\dagger \hat{c}_j,
    \label{HF}
    \ea
    where \(h_0\) is the hopping amplitude and \(\mu\) the chemical potential. We probe the density-wave order \(\hat{\rho}_q = \frac{1}{L}\sum_k \hat{c}_{k+q}^\dagger \hat{c}_k\) under dissipation on even sites ($L$ is the system size). The  DS for $\hat{W}=\hat{\rho}_{q}$ and  \(\hat{O}_j = \hat{n}_j\) for all even sites is defined as $\chi^{\mathcal{D},[0]\Re}_{e,\hat{\rho}_q}=\sum_{j\in 2\mathbb{Z}}\chi_{\hat{n}_j,\hat{\rho}_q}^{\cal D}(\omega)$, explicitly
    \ba
    \chi^{\mathcal{D},[0]\Re}_{e,\hat{\rho}_q}\!(\omega)\!=\! \delta_{q,\pi} \!\!\sum_k\! \!\left[ 2\bar{n}\!-\! n_{k\!+\!\frac{\pi}{2}}\!-\! n_{k\!-\!\frac{\pi}{2}} \right] \!\!\frac{\delta(\omega \!+\!\Delta_{{\pi}}\sin k)}{2L}\!,
    \ea
    where $\Delta_q = 4h_0\sin\frac{q}{2}$, $\bar{n}$ is average density, and $n_k$ the Fermi-Dirac distribution. We numerically validate the measurement protocol on a 10-site chain (6 particles) with periodic boundary condition(PBC) in the zero-temperature limit, monitoring the even–odd particle imbalance \( (N_e \!-\! N_o)(t) = L\langle \hat{\rho}_\pi(t) \rangle\). Dissipative dynamics are computed via the Lindblad master equation. Fig 1(b) shows linear response regime for various \(\gamma\) ($\gamma/h_0=0.005$, $0.01$ and $0.02$). With an additional modulated \(\gamma(t) = \gamma + \gamma'\cos(\omega t)\), the extracted response amplitude scales linearly with \(t\) (Fig. 1(c)), and the reconstructed DS agrees closely with finite-size theoretic result (Fig. 1(d)). This confirms the reliability of the proposed dynamical scheme for DS. Furthermore, Fig 1(b) shows exponential-decay tendency. By multiplying $e^{\alpha \gamma t}$ with a proper $\alpha$, the linear order contribution can be recovered \cite{supplementary}. By this discovery, we can work in $\omega_0^{-1}<(\gamma/10)^{-1}$ region with an enlarged range for dissipation strength.
    \begin{figure}[t]
       \includegraphics[width=8.5cm]{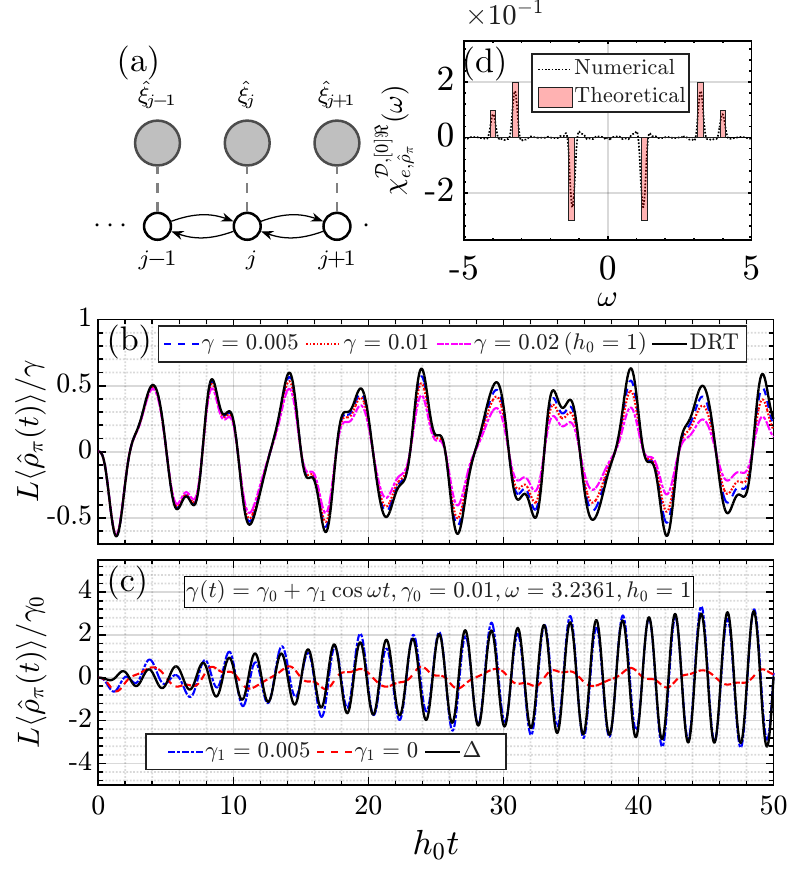}
            \caption{(a) Free-fermion chain locally coupled to a dissipative bath. (b) Dissipation dynamics of the even-odd particle imbalance $N_e - N_o$ for varying dissipation strengths $\gamma$. (c) Applying an extra oscillation in dissipation strength, a resonance behavior is found in the even-odd particle number deviation. (d) Dissipative spectroscopy extracted from the resonance signal in (c). Our protocol agrees well with theoretic result.}
        \end{figure}

        \begin{figure}[b]
            \centering
            \includegraphics[width=8cm]{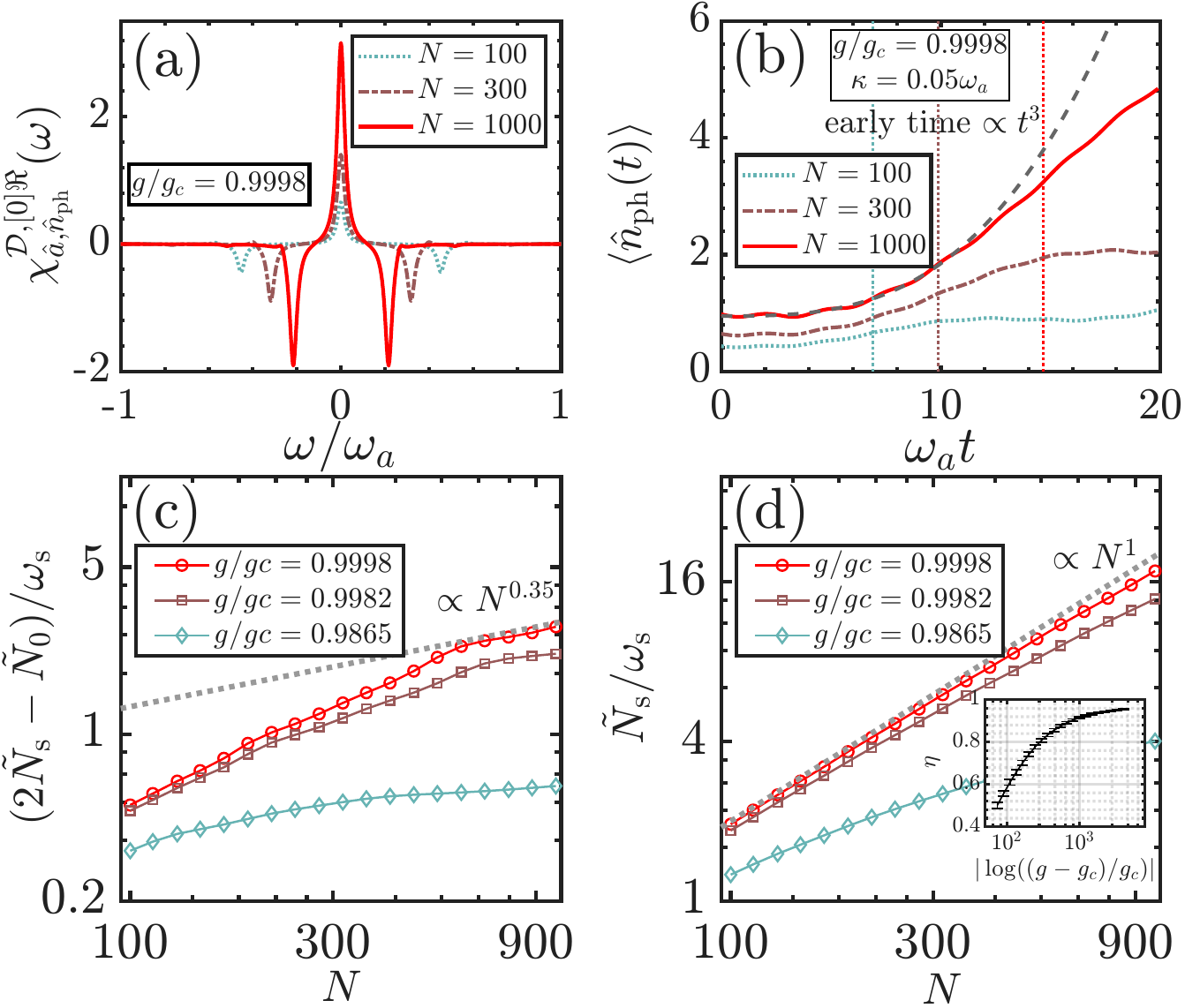}
            \caption{(a) Dissipative spectrum $\chi^{\mathcal{D},[0]\Re}_{\hat{a},\hat{n}_{\rm ph}}(\omega)$ for $g/g_c=0.9998$ at different $N$. (b) Photon number dynamics under dissipation ($\kappa=0.05\omega_a$). For $N\!=\!1000$, early-time growth follows $t^3$. (c) Scaling of $(2\tilde{N}_{\rm s}\!-\! \tilde{N}_0)/\omega_{\rm s}$ with $N$ near criticality. This quantity is smaller than $N^{0.35}$ for large $N$. (d) $\tilde{N}_{\rm s} /\omega_{\rm s}$ as a function of $N$. Log-log plot reveals $\tilde{N}_{\rm s} /\omega_{\rm s} \propto N^{\eta}$ in the critical region, $\eta\approx0.95$ for $g=0.9998g_c$.}
            \label{fig:F2}
        \end{figure}
        \emph{\color{blue}Dissipative quench quantum criticality\color{black}}. --- To demonstrate the capability of dissipative spectroscopy, we study the dissipative quench dynamics of the Dicke model \cite{Dicke54}. We show that the scaling behavior extracted from the DS reveals a nontrivial \(t^3\) growth of the cavity photon number in the normal phase, leading to an approximately macroscopic photon number occupation that scales extensively with the system size (as $N^{0.95}$). This observation provides insight into the unresolved discrepancy between theoretical and experimental critical exponents in superradiant transitions \cite{Esslinger12,Esslinger15,Hemmerich14,Brantut25,Sachdev13,Domokos12,EsslingerRev13,RitschRev22}, where dynamical dissipation effects have not been fully accounted for. 
        
        Specifically, we consider a Dicke Hamiltonian:
        \ba
        H= \omega_c\, \hat{a}^{\dagger} \hat{a} + \omega_a \hat{J}_z + \frac{2g}{\sqrt{N}}(\hat{a}^{\dagger} + \hat{a}) \hat{J}_x,
        \ea
        where \(\hat{a}\) is the cavity photon annihilation operator, \(\omega_c\) (\(\omega_a\)) is the cavity (atomic) frequency, and \(\hat{J}_\alpha = \frac12 \sum_{i=1}^{N} \hat{\sigma}_\alpha^{(i)}\) are collective spin operators ($i$ labels the spins). Here, $N=2J$ denotes the total spin. The superradiant phase transition occurs at \(g_c = \tfrac12\sqrt{\omega_a\omega_c}\) for $\kappa=0$ in the $N\rightarrow\infty$ limit. Here, we avoid reaching thermodynamic limit, studying a finite-$N$ system.
        
        We implement a dissipative quench by suddenly turning on photon loss at rate \(\kappa\) and measuring the photon number \(\hat{n}_{\text{ph}} = \hat{a}^\dagger \hat{a}\), setting $\omega_c = 2\omega_a$. The DS is defined as \(
        \chi^{\mathcal{D},[0]\Re}_{\hat{a},\hat{n}_{\rm ph}}(\omega)= 2\int_{-\infty}^\infty dt \, \langle\hat{a}^\dagger\hat{n}_{\rm ph}(t)\hat{a}-\frac{1}{2}\{\hat{a}^\dagger\hat{a},\hat{n}_{\rm ph}(t) \} \rangle \, e^{i\omega t}\), where $\langle\cdots\rangle$ denotes the ground-state average. By exact diagonalization, we obtain
        \ba
        \chi^{\mathcal{D},[0]\Re}_{\hat{a},\hat{n}_{\rm ph}}(\omega) \approx\tilde{N}_0\delta(\omega)-\tilde{N}_{\rm s}\sum_{s=\pm}\delta(\omega+s\omega_{\rm s})+\cdots,
        \label{chi}
        \ea
        where  \(\omega_s\) is the soft-mode frequency extracted from DS\cite{supplementary}, \(\tilde{N}_s\) is the associated spectral weight, and \(\tilde{N}_0\) is the zero-frequency spectral weight.  From Eq.~\ref{chi}, the early-time dissipative dynamics follows
        \ba
        \langle\hat{n}_{\rm ph}(t) \rangle \!\approx \!\frac{(\tilde{N}_0 \!-\! 2\tilde{N}_{\rm s})}{\omega_{\rm s}}\omega_{\rm s}t \!+\!\frac{\tilde{N}_{\rm s} }{3\omega_{\rm s}} (\omega_{\rm s}t)^3 \!+\! \mathcal{O}(t^5).
        \ea
        Near the critical coupling $g_c$ from the normal side, we find the scaling relations $\omega_{\rm s}\propto N^{-\nu_1}$, $\tilde{N}_0\propto N^{\nu_2}$, and $\tilde{N}_{\rm s}\propto N^{\nu_3}$. Exact diagonalization at $g=0.9998g_c$ yields
        \ba
        \nu_1\sim 0.33,\quad\nu_2\sim 0.68,\quad \nu_3\sim 0.62.
        \ea
        The $\langle \hat{n}_{\rm ph}\rangle$ reaches $\tilde{N}_{\rm s}/\omega_{\rm s}$ at $t\sim\omega_{\rm s}^{-1}$, giving
        \ba
        n_{\rm ph}(\omega_{\rm s}^{-1})\sim\tilde{N}_{\rm s}/\omega_{\rm s}\propto N^{\nu_1+\nu_3\approx 0.95}.
        \ea
        The critical dissipative quench dynamics is shown in Fig.~\ref{fig:F2}(b). We emphasize that the photon number can be regarded as macroscopically occupied for two reasons. First, its scaling is nearly linear in $N$. Second, the absolute number is substantial: for $10^3$ atoms we obtain about 5 photons, compared to about 50 photons from $10^5$ atoms in previous Dicke transition experiment {\cite{Esslinger10}}.
        
        We now address the finite-size nature of our analysis. Both $\tilde{N}_{0}$ and $\tilde{N}_{\rm s}$ obey a universal finite-size scaling of the form $\mathcal{A} = \mathcal{A}_0[1-e^{-(N/\ell_c)^\beta}]$\cite{supplementary}, where the critical size $\ell_c \propto (g_c-g)^{-\nu}$ with $\nu \sim 1.5$, and $\beta=\nu_2$ ($\nu_3$) for ${\cal A}=\tilde{N}_0$ ($\tilde{N}_{\rm s}$). This reveals two regimes: a finite-size critical region for $N<\ell_c$ {\cite{VidalDusuel,DusuelVidal2004,Chen08}}, and a near-thermodynamic-limit regime for $N>\ell_c$. If the thermodynamic limit is taken first, only the latter regime is visible. However, many experiments operate in the finite-size regime, where critical behavior manifests differently. The DS thus provides a meaningful probe of finite-size criticality, complementary to the conventional thermodynamic-limit description.

        \emph{\color{blue}Memory effect described by generalized dissipative susceptibilities\color{blue}}. --- To test the validity of dissipative response functions as memory kernels beyond the Markovian regime, we study a one-dimensional fermionic chain locally coupled to an ensemble of SYK$_2$ quantum dots providing structured baths with finite temporal correlations. A quench at $t=0$ is implemented by switching on the system–environment coupling at even sites.
        
        The total Hamiltonian reads
        \ba
        \hat H_{\rm tot}=\hat H_0+\hat V,\qquad 
        \hat H_0=\hat H_{\rm FF}+\hat H_{\rm SYK_2},
        \ea
        with \(\hat H_{\rm FF}\) given in Eq.~\ref{HF},
        \(\hat H_{\rm SYK_2}=\sum_{i,\alpha\beta}J^i_{\alpha\beta}\hat\psi^\dagger_{i\alpha}\hat\psi_{i\beta}\), describing the environment, 
        \(\hat V=\sum_{i\in\Lambda_e,\alpha\beta}V^i_{\alpha\beta}\hat\psi^\dagger_{i\alpha}\hat\psi_{i\beta}\hat n_i\), the system-environment coupling.
        Here \(\hat\psi_{i\alpha}\) are bath fermions (\(\alpha=1,\dots,M\)). The random couplings \(J^i_{\alpha\beta}\) (\(V^i_{\alpha\beta}\)) have zero mean and variances \(J^2/M\)  (\(V^2/M^2\)). We consider a  10-site chain (6 particles) under PBC, with the system prepared at inverse temperature \(\beta_{\rm S} h_0=1\) and the environment at temperature \(T_{\rm E}=0\). Within DRT, \(\hat O_i=\hat n_i\) and \(\hat\xi_i=\sum_{\alpha\beta}V^i_{\alpha\beta}\hat\psi^\dagger_{i\alpha}\hat\psi_{i\beta}\).
        In the large-\(M\) limit, the environment is assumed to experience negligible backaction from the coupling.

        To assess the accuracy of the non-Markovian DRT, we require a reliable dynamical solution for physical observables. We therefore employ  Keldysh formalism and derive the Kadanoff–Baym equation (KBE){\cite{BaymKadanoff1961,Baym1962}} for the greater/lesser two-point Green’s functions
        \(
        G^>_{ij}(t_1,t_2)\!=\!-i\langle \hat{c}_i(t_1) \hat{c}^{\dagger}_j(t_2)\rangle ,\quad
        G^<_{ij}(t_1,t_2)\!=\!+i\langle \hat{c}^{\dagger}_j(t_2)\hat{c}_i^{}(t_1) \rangle
        \) :
        \begin{align}
            (i\partial_{t}-\hat{H}_0)\circ G^\gtrless&=\Sigma^R\circ G^\gtrless+\Sigma^\gtrless\circ G^A\nonumber\\
            G^\gtrless\circ(i\partial_{t}-\hat{H}_0)&=G^R\circ \Sigma^{\gtrless}+G^\gtrless\circ \Sigma^A.
        \end{align}
        Here the convolution is defined as $A\circ B(x,x'')=\int dx' A(x,x')B(x',x'')$ (Here $x$ includes space and time index), and $G^{R}=\theta(t_{12})(G^>-G^<)$, $G^{A}=\theta(t_{21})(G^<-G^>)$. The greater/lesser self-energies are
        \ba
        \Sigma^{\gtrless}_{ij}(t_1\!,\!t_2)\!\!&=&  \!\!\delta_{ij} \delta_{i \in 2\mathbb{Z}}\! \Big[ g^{\gtrless}_\xi(t_1,t_2) G^{\gtrless}_{ij}(t_1,t_2)\!- \nonumber\\
        \!\! &&\! \!2i\delta(t_1\!-\!t_2) \!\!\int_{0}^{t_1} \!\!\!\!\!d\tau \, \Im (g_\xi(t_1,\tau)) G^{\gtrless}_{ij}(\tau,\tau) \!\Big]\!,
        \ea
        with $g^>_\xi(t_1,t_2)=g^{}_\xi(t_1,t_2)\equiv V^2 G_\psi^>(t_1,t_2)G^<_\psi(t_2,t_1)$ and $g^<_\xi(t_1,t_2)=g^{}_\xi(t_2,t_1)$. The SYK$_2$ fermion propagators $G^\gtrless_\psi(t_1,t_2)$ are known analytically \cite{supplementary}. The corresponding Feynman diagrams are shown in Fig. 3(a).
        \begin{figure}[t]
            \centering
            \hspace{-3ex}
            \begin{overpic}[width=7.2cm]{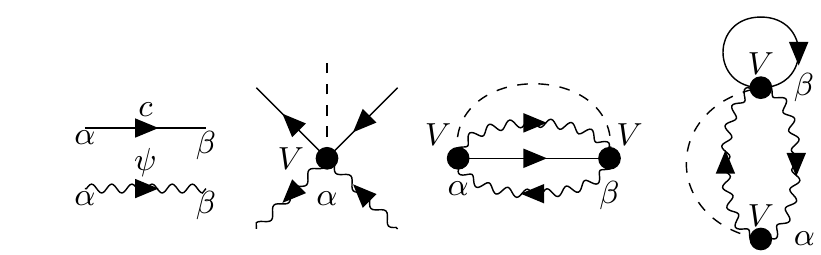}
                \put(21,22){\normalsize \textcolor{black}{\text{(a)}}} 
                \centering
            \end{overpic} \newline
            \\
            \begin{overpic}[width=7.0cm]{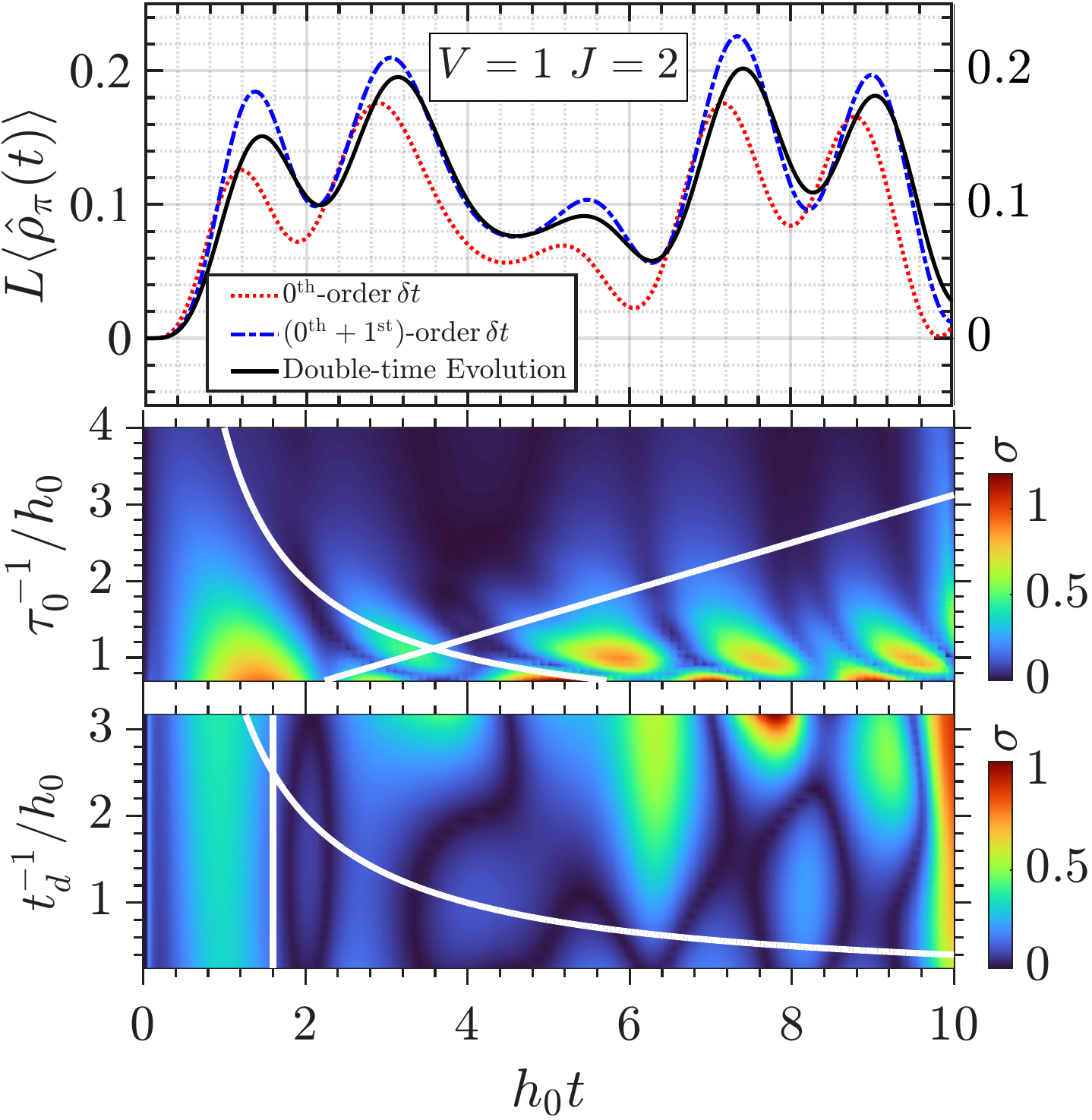}
                \put(15,95){\normalsize \textcolor{black}{\text{(b)}}} 
                \put(15,57){\normalsize \textcolor{white}{\text{(c)}}} 
                \put(15,31){\normalsize  \textcolor{white}{\text{(d)}}} 
            \end{overpic} 
            \caption{(a) Feynman diagrams for fermions $c$ and $\psi$.  Disorder average is indicated by dashed line; density-density interaction appears as vertex labeled by $V$. Two types of self-energies from system-bath coupling are shown. (b) Dissipation dynamics of the even-odd imbalance $N_e - N_o$: numerical result based on KBE (black solid), zeroth-order DRT (red dotted), and combined $0+1$-order response (blue dash-dotted). (c, d) Variance $\sigma(t)$ comparing full evolution to the leading-memory response theory. (c) Fixed $V=1$, varying memory time $\tau_{0} = 1/J$; (d) fixed $\tau_{0}=1/2$, varying dissipation time scale $t_d=J/V^2$. White lines mark empirical boundaries $\propto \tau_{0}$ and $t_d$, which the low variance confirms the validity of DRT.}
        \end{figure}

        Using a standard predictor–corrector time-stepping algorithm \cite{SachdevB2017,Altman2017,Ryu2022,Chen2022}, we evolve $G^\gtrless(x_1,x_2)$ in time, from which the local density $-iG^<_{ii}(t,t)$ is obtained. The resulting evolution of the site-imbalance $(N_e-N_o)(t)$, computed via this double-time KBE approach, is plotted as the solid black curve in Fig. 3(b).
        
        Within the DRT framework, the leading-order dissipative dynamics is given by
        \ba
        \delta\mathcal{W}(t)\approx \!\!\sum_{\ell,j\in\Lambda_e}\!\!\left[ \chi^{\mathcal{D},[\ell]\Re}_{\hat{O}_j,\hat{W}} \!\circ\! {\cal G}^{>,[\ell]\Re}_\xi \!\!+\chi^{\mathcal{D},[\ell]\Im}_{\hat{O}_j,\hat{W}} \!\circ\! {\cal G}^{>,[\ell]\Im}_\xi\right]\!\!(t),
        \ea
        where $\ell=0,1$, and $\chi^{\cal D}$ and ${\cal G}_\xi$ are defined in Eqs.~\eqref{def_g}–\eqref{def_chi_2}. Notably, ${\cal G}_{\xi}^{[0]\Re}(t)\to0$ as $t\to \infty$ at zero temperature, implying that the Markovian-limit response vanishes. Hence, all dissipative dynamics in ${\cal W}=N_e-N_o$ arise from memory effects. The zeroth order contribution and the $0+1$ order contributions are shown in Fig.~3(b).
        
        For $V=1$ and $J=2$, the  $0+1$ order memory corrections describe the dynamics accurately. To quantify the consistency, we define a time-dependent deviation
        \ba
        \sigma(t)=\frac{\int_{t-\delta_0}^{t+\delta_0}d t' \, \bigl(\Delta N^{\text{KBE}}(t')-\Delta N^{\text{DRT}}(t')\bigr)^2 }{\int_{t-\delta_0}^{t+\delta_0} d t' \, \bigl(\Delta N^{\text{KBE}}(t')\bigr)^2},
        \ea
        where $\Delta N^{\rm KBE}$ and $\Delta N^{\rm DRT}$ denote the imbalance obtained from the KBE and DRT calculations, respectively, and $\delta_0$ is a coarse-graining window taken as $(5h_0)^{-1}$. As shown in Fig. 3(c) and (d), $\sigma(t)$ remains negligible for times $\tau_0<t<t_d$, confirming that the generalized dissipative susceptibilities capture the memory effects accurately within this interval. Here $\tau_0\equiv 1/J$ is the memory time defined by the correlation time in $g_\xi(t_1,t_2)$. $t_d\equiv J/V^2$ is the dissipation time scale, whose inverse describes the dissipation strength. The same protocol can be applied at high temperature, where the short correlation time $1/J$ validates the Markovian approximation.
        
        \color{blue}\emph{Conclusion}\color{black}. --- Here we introduce a definition of dissipative spectroscopy. We design a protocol to probe the DS dynamically in a dissipation-driven resonance. Then we find that the DS may reveal finite size criticality (in contrast to thermal dynamical limit criticality) in a dissipation quench at normal phase side. Finally, we find generalized dissipative susceptibilities describing the memory effects in dissipative dynamics well in a time-scale in between the memory time and the dissipation time. 
        
        \color{blue}\emph{Aknowledgement}\color{black}. --- This work is supported by the National Natural Science Foundation of China (Grants No. 12174358), the National Key R$\&$D Program of China (Grant No. 2022YFA1405302), and NSAF (Grant No. U2330401).

\end{document}